# EXPERIENCE BUILDING A 5G TESTBED PLATFORM

Mona Ghassemian
BT Applied Research
Adastral Park, Ipswich, IP5 3RE, UK
mona.ghassemian@bt.com

Paul Muschamp
BT Applied Research
Adastral Park, Ipswich, IP5 3RE, UK
paul.muschamp@bt.com

Dan Warren
Samsung R&D Institute
Staines-upon-Thames, TW18 4QE, UK
dan.warren@samsung.com

*Abstract*— The 5G-VINNI testbed infrastructure project provides 5G facilities for pan-European services. Within the project, the UK site is one of the 5G-VINNI facilities that targets developing a flexible and dynamic test environment which can be adapted to meet requirements from H2020 funded projects as well as external trials, to enable vertical industries to assess 5G networks in the context of advanced digital use cases. In this paper, we present a full overview of the 5G test infrastructure developed at the UK 5G-VINNI facility, including backhaul, edge, slicing, interworking and validation. This work presents an operator's experience in setting up a testbed facility and in engaging with different verticals conducting their research projects using the 5G facility, and the results and benefits to the industry. Furthermore, the lessons learned from the design, installation, operation and experimentation phases of the project are discussed.

## I. INTRODUCTION

5G is currently being deployed around the world, providing mobile users with an enhanced service compared to previous generations of cellular networks. At the same time, a number of 5G testbeds and trial networks are being deployed that allow the industry to leverage the enhanced capabilities of the new networks, beyond simply a 'faster version of 4G'.

Horizon 2020 is Europe's primary research and innovation programme. It supports the development, evolution and roll-out of 5G networks, through the formation of consortia made up of equipment vendors, operators, small-medium enterprises and academia. The programme is overseen by the 5G Infrastructure Public Private Partnership (5G PPP). The system makes use of a series of funding calls (such as ICT calls) aimed at pursuing a particular aspect of 5G, including the development of a series of inter-connected 5G testbeds around Europe. The aim of the testbeds is to i) demonstrate that the key 5G PPP network KPIs can be met; ii) be validated, accessed and used by vertical industries to set up research trials of innovative use cases, to further validate core 5G KPIs in the context of concurrent usages by multiple users.

Three projects have been funded by H2020 ICT-17-2018 - 5G-EVE, 5GENESIS, and 5G-VINNI [1]. The three platforms that these projects have developed enable several vertical use cases from dedicated 5G vertical industry projects [2] to be demonstrated and validated. The platforms are also open to other 5G trials from industry [3] and academic institutes, to conduct research for emerging 5G ideas, proving concepts, validating standards and vendor interoperability testing [4].

In this paper, we present 5G-VINNI (5G Verticals Innovation Infrastructure) [1]-[6] which provides a set of inter-connected testbeds with main sites (aka. facility) in UK, Norway, Spain and Greece [5] as shown in Fig. 1. Each end-to-end network facility covers fixed/multi radio access, backhaul, core network, service technologies and architectures targeted for 5G including end-to-end virtualisation and slicing as key components to support vertical use cases.

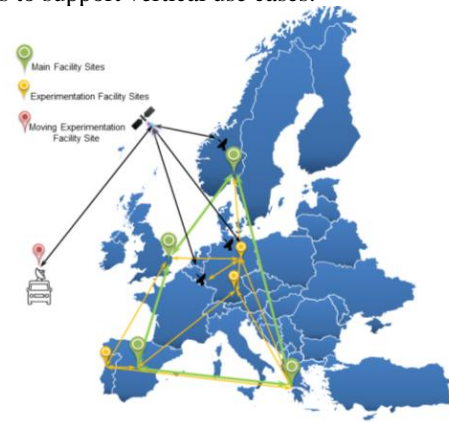

Fig. 1.  *5G-VINNI infrustructure cartography*

The 5G-VINNI UK facility _which is the focus of this paper_ is deployed at Adastral Park, BT's main research and development centre in UK. The UK facility provides a 3GPP-compliant, end-to-end 5G system, with virtualised RAN and core, and is deployed to support vertical industries based in the UK and Europe to test their 5G applications. This paper provides a description of the UK facility, verified for a set of use case experiments – the aim being to show the 'art of the possible' and sign-post the direction in which 5G will evolve.

The paper is organised as following: Section II provides an overview of 5G systems and industry outlook. Section III presents an end-to-end overview of the UK facility and target capabilities, and shares deployment experience gained from setting up the platform. Section IV outlines use cases and applications that are being validated using the UK 5G-VINNI facility. Section V shares the experience of setting up the facility site and lessons learned. Finally, Section VI concludes the paper.

## II. 5G OVERVIEW & OUTLOOK

The launch of 5G commercial networks during 2019 was the culmination of years of research, standardisation and product development. However, 5G still has some way to go before it can deliver all of the potential benefits, including boosts to performance, higher revenue by increased efficiency, automation, orchestration and scalability, as the result of using virtual and softwarised solutions, that are the objectives of the telecoms industry.

In the NGMN 5G Whitepaper published in 2015 [7], requirements are gathered on the basis of commercial imperative to address new market segments. The opportunities include healthcare, Industrial Internet of Things (IIoT) and Industry 4.0, Smart Cities, Vehicle-to-anything (V2X) connectivity, and agricultural applications. From this list, a set of requirements was derived, each presented as a quality metric relating to data rate, latency, connection density, minimum data rate available in any location, coverage, reliability and availability, and Total Cost of Ownership (TCO). The metrology of most of these is well understood, but TCO requires considerable evaluation. At the top level this is the sum total of Capital Expenditure (the costs associated with the establishment of the network) and Operational Expenditure (the costs associated with operating the network).

Capital Expenditure (CapEx) includes the cost of equipment for the network and maintaining the network integrity (back-up power supply, air conditioning, etc.), spectrum costs, civil engineering materials and costs, connectivity infrastructure from front and backhaul, site purchase, rights of way and planning permission. Therefore, CapEx is impacted directly by achieving other metrics included in the assessment of 5G. For example, to improve network reliability, equipment redundancy and route redundancy need to be higher than for a less reliable operation. This then requires more equipment and redundancy in routing to be included in Capital costs. Similarly, to achieve high levels of coverage, radio equipment needs to be installed in places where previously networks may not have been economically viable, such as provision of indoor coverage, or in remote rural areas.

Operational Expenditure (OpEx) includes network powering, leased line connection costs, rent of sites, staff costs, and maintenance costs. The increased levels of equipment needed for reliability and coverage result in a potential increase in operational costs since there is more equipment to be housed, so more sites need to be connected, powered and maintained.

GSMA's 5G Whitepaper [8] reflects that while 5G is at its root a set of technological advancements, meeting requirements such as increased reliability, availability and network coverage is only possible with investment in building and maintaining a 5G network which has these metrics incorporated in its design

Therefore, 3GPP Release 15, which forms the basis of all commercial 5G network launches to date, cannot address all of the industry requirements that 5G is expected to support within its specifications. While commercial 5G networks do deliver very high data rates and a significant reduction in latency, few consistently meet headline rates of 1Gbps or target latency of below 10ms. Unsurprisingly, none offer 100% coverage as yet, and network reliability and availability can only be addressed over the long term, once the network is stressed under high subscriber numbers.

5G commercial networks to date are primarily focussed on the consumer market. Early 5G devices are generally high-end smartphones, while Fixed Wireless Access (FWA) CPE and 5G USB modems are also available. Very little has yet to be launched in terms of a specialised commercial 5G B2B offering.

The reason for this is multi-faceted: First, 3GPP Release 15 includes two deployment models for 5G RAN – Non-StandAlone (NSA) and StandAlone (SA). NSA was included in R15 as a quick fix to allow commercial implementation of 5G New Radio (NR) to begin earlier than might have otherwise occurred. NSA requires simultaneous attachment to both 5G-NR RAN and 4G LTE RAN air interfaces and connects back to the 4G Evolved Packet Core (EPC). 5G SA connects directly to the 5G Core, which is defined as a Service-Based Architecture (SBA), as described in 3GPP TS 23.501 [10]. It is widely acknowledged that there is incompatibility between NSA and SA-based networks and devices. With many B2B opportunities being dependent upon a time-of-life for equipment that is far longer than that of commercial smartphones, it is understandable that B2B customers are reticent to deploy NSA, which is recognised to be a relatively short-term deployment option.

Second, the production scale of 5G radio modules is still relatively low, and the technological advancement in 5G is high, meaning that 5G component costs for devices represents a premium that is not justifiable for most B2B implementations.

In defining 5G, many requirements have been pushed to extremes. The requirement for 1Gbps connections has led to extensions into massive MIMO, as well as driving work on beam forming to enable the use of frequency bands far above 6 GHz (aka mmWave, although many of these bands are not strictly in the mm wavelength domain). 1ms latency is only achievable if backhaul links are kept short, which can be done with Multi-access Edge Computing (MEC) hosted User Plane Function (UPF). Where backhaul links are long, the speed of light (particularly when using glass fibre) becomes a limiting factor.

Many of the use cases that are identified for 5G do not require all of the most challenging key performance indicators (KPIs) of 5G to be supported at once. This means the dedicated network instances supporting different network topologies, each enabling a specific class of service, can be implemented. This is achieved by the definition of 'network slicing' – the support of multiple, virtualised network instance, logically separated on common hardware and infrastructure. Network slicing is a key 5G service enabler, but the practicality of implementing individual network instances for different services, all sharing common infrastructure and resource is in itself a considerable new challenge. It has led to the need for the Management and Network Orchestration (MANO) domain to be defined. MANO and slicing have then resulted in the definition of Slice Lifecycle Management.

The industry is thus, still discovering a lot about what 5G is, what it might be in the future, and what the use cases are that it can enable that will propel it forward and compel operators to support increasingly challenge network topologies and management requirements. With the vast majority of 5G connections still being for B2C customers, and using traditional business models of subscription and/or usage based charging, 5G today is little more than an extension of 4G service. For 5G to meet the promises it has made, further research and development to truly stress the network's capabilities is required. Through the implementation of the UK 5G-VINNI test facility, it is intended to support this research and experimentation on both the network side, to push network capabilities to their limits, and to offer a new perspective to other industries about what 5G will enable in the mid- to long-term.

### III. UK 5G VINNI FACILITY

5G-VINNI project deploys multiple testbed facilities, to enable future projects and outside parties to experiment with 5G use cases. The 5G-VINNI testbed platform developed at Adastral Park in the UK, has been enabled using Samsung network equipment, and is implemented in two phases.

The first implementation phase, completed in August 2019, provides two different network types. The first is a 3GPP compliant 5G-NR radio in the 3.6 GHz band (Band n78 TDD), operating in NSA mode in conjunction with an LTE RAN in Band 1, compliant with 3GPP TS 38.401 [9]. The 5G-NR functional components are implemented using Radio Unit – Distributed Unit – Centralised Unit split architecture (RU-DU-CU split), with the DU and CU components being virtualised. In this configuration, the CU hosts Radio Resource Control (RRC) and Packet Data Convergence Protocol (PDCP); The DU hosts the Radio Link Control (RLC), the Medium Access Control (MAC) layer and the higher parts of the physical (PHY) layer; and the RU handles the digital front end and lower parts of the PHY layer. This in turn means that the architecture exposes a Common Public Radio Interface (CPRI) between RU and DU, and the F1 interface between CU and DU as shown in Fig. 2. This amounts to implementation of Option 2 and Option 7 of the functional decomposition of the RAN as described in 3GPP TS 38.801 [11].

The RAN elements connect back to a virtualised EPC (vEPC). The vEPC is implemented as Virtual Network Functions (VNFs) with in a Network Function Virtualisation Infrastructure (NFVI) architecture. The NFVI architecture itself is split into three domains – the Compute and Storage domain, made up of Common Off-The-Shelf (COTS) infrastructure; the Hypervisor Domain, which is implemented using Red Hat Enterprise Linux (RHEL) 7.x as the host OS and Kernel-based Virtual Machine (KVM) as the hypervisor; and the Infrastructure Network Domain implemented using BT's in-house transport network between physical nodes and using rack backplane for intra-rack, inter-server connections. The network is orchestrated from an ETSI compliant Management and Network Orchestration (MANO) architecture, compromised of Virtualisation Infrastructure Management (VIM), VNF Managers (VNFM) and NFV Orchestrator (NFVO).

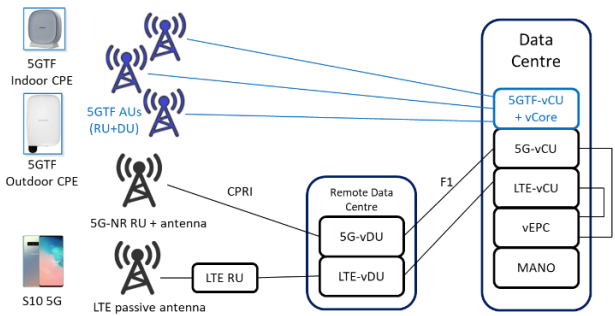

Fig. 2. *Top level architecture*

Separate to this, the second network implementation in phase 1, is a 5GTF-based mmWave (26GHz band) implementation offering Fixed-Wireless access (FWA) connectivity to a choice of indoor and outdoor customer premise equipment (CPE). The 26 GHz 5GTF air interface is provided by an integrated radio Access Unit (AU), comprised of radio unit (RU) and DU, which connects directly to its own 5GTF CU and core. CU and Core are also fully virtualised but are implemented statically with no orchestration architecture.

The physical implementation of antenna currently is comprised of a single sector of 5G-NR NSA, and three 5GTF AUs. Locations for these antennas on the Adastral Park site has been chosen to provide full site coverage for both radio systems. The site is a business park with buildings, roadways and pedestrian walkways. As is described in Section IV, a wide variety of use cases demand that the radio provides coverage for vehicles, pedestrians and building occupants. Antenna locations have been chosen based upon available & accessible roof spaces and on those that provide the coverage desired. Fig. 3 shows the 3.6 GHz & LTE antennas installations in the left-most photograph and the mmWave antennas in the other three.

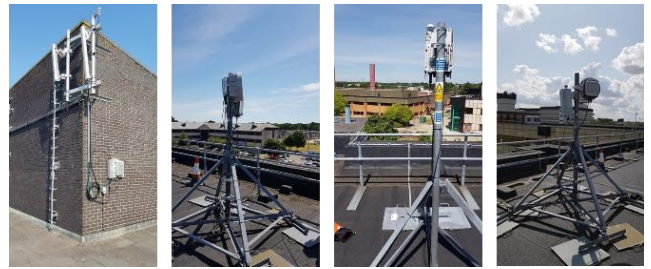

Fig. 3. 5G-VINNI antenna installations at Adastral park

This first phase of installation provides a stable, production-equipment based platform that can be used to support multiple experiment types, delivering high bandwidth and low latency. For the 5G-NR connection, commercial Samsung Galaxy S10 devices are used, whilst for 5GTF, connection from CPE to experimental devices is via Ethernet cable.

In the second phase, expected to be implemented during the second half of 2020, the RAN networks in both 3.6GHz and 26GHz bands will be moved to 5G-NR SA, connecting to a common SBA-based core. This will be controlled by the virtualised MANO domain with enhanced support for network slicing, and the possibility of including network topology flexibility options through the introduction of Multi-Access Edge Computing (MEC) capability.

### IV. UK TEST PLATFORM SET UP

5G-VINNI facilities are set up to enable a flexible and dynamic testbed environment which can be adapted to meet the requirements of vertical trials. Experimentations concentrate on the enhanced mobile broadband (eMBB) and ultra-reliable low latency communications (URLLC) capabilities first and massive machine-type communications (mMTC) when demanded. We provide an overview of the current testbed set up on the 5G-VINNI UK infrastructure. In each case, the use case for each potential experiment is described in terms of the required network supports, the verification for the suitability of the set up and benefits to the industry. A summary of the experimentations is listed in TABLE I.

*A. Cloud-based gaming*

A major boost to cloud-based multi-player gaming is expected from its current user base of PC/games consoles to mobile devices, as 5G provides the required eMBB and uRLLC capabilities.

In an initial verification on the 5G-VINNI UK facility, following the network commissioning in 2019, we made an empirical comparison of the performance of the UK 5G-VINNI facility compared to 4G and WiFi/broadband networks for indoor and outdoor scenarios using a typical multi-player car racing game.

In the outdoor site verification, we compared a cloud-based gaming application running on a Samsung Galaxy S10 connected to commercial 4G service, and the 3.6 GHz 5G NR of the 5G-VINNI UK facility. We observed that while it was possible to access the cloud to load the game over 4G, when attempting to play the game, it frequently dropped out or stopped working altogether. However, 5G connection allowed the user to play the game with acceptable latency for control change of racing car movement, and with reasonable resolution on the S10 handset.

In the indoor site verification, we compared the same cloud-based racing game on a high-end gaming laptop connected to a local WiFi access point connected to BT's internal network, with the 26 GHz radio of the 5G-VINNI UK facility. Our observation showed that the game loaded and ran well using WiFi but suffered from intermittent latency-based command control problems, (e.g., a signal being delayed making the required change in the car's direction, resulting in a crash) or periods of poor pixilation of the game display. However, the game ran well using 26 GHz connection with high control response and display resolution.

Although this experiment provides only empirical and subjective evidence of the capability of the 5G-VINNI network in a cloud gaming environment, the results demonstrated a key 5G technology role in cloud gaming for the mobile game market.

*B. Assisted living*

Healthcare is increasingly making use of technology to complement the efforts of care workers, and 5G has its part to play. Considering the growing aging demographic, assisted living can transform the lives of millions of people in the UK alone. People in assisted living residentials are keen to retain as much independence as possible but want the reassurance that they are safe and secure [12].

The use of ambient monitoring sensors and wearables connected over 5G-VINNI help collect information particularly from vulnerable people, and help build behavioural patterns based on their daily routine model. In case of an accident, an AI system can alert care workers based on the anomaly detection, turn on a video camera and use the 5G-VINNI network to stream video to a care worker's 5G handset. The collected data is used to build a set of connected care dashboards for assisted living operations. Furthermore, we have applied video analytics to support assisted living in which data from cameras to determine whether situations demand help for care workers. This includes the use of drones connected over 5G-VINNI to spot when people in assisted living have wandered away from the care home.

*C. Remote robotic control and immersive telepresence*

Network slicing has been touted as a key 5G differentiator, offering the ability to offer discriminating levels of service to customers who demanded a premium service, particularly by Industry4.0 use cases. In advance of the 5G-VINNI network providing network slicing, it can already support the provision of simultaneous paths through the network, supporting both eMBB and URLLC capabilities. Using these paths, a robotic control application over 5G-VINNI showed tens of milliseconds latency for remote control of the robot, as well as over 100s of Mbps of bandwidth for streaming 360° video content to a virtual reality (VR) headset.

In this experiment, we have combined the two, using the low latency path to undertake remote control of a movable robot and using the high bandwidth path to provide the control user with an immersive experience of seeing a 360° view from the robot's cameras. The user makes use of a game controller which allows forward, backward, sideways, and spinning controls to perform quite intricate control of the movable robot [13].

*D. Dedicated QoS supporting outside broadcast media*

Outside broadcast is a key part of news and sports coverage where the camera needs to be away from the studio. Rapid-response media in mobile ad hoc situations demands a high-performance network able to support a high bandwidth path and processing capability that ensures quality of service (QoS).

The use of 5G-VINNI for remote broadcast production must provide up to 1Gbps for downstream content delivery and up to 250Mbps for upstream content delivery. It must also be suitable for mezzanine-compressed HD video or more heavily compressed UHD video, in which the network provides a bandwidth and packet loss performance guarantee.

This experiment uses the 5G-VINNI 26 GHz radio to demonstrate that the bandwidth performance satisfies the needs of the broadcast application, but to reduce the upstream packet loss customised network slicing needs to be implemented.

*E. Low-latency situation recognition*

Many public safety and industrial applications require the use of situation recognition using cameras to detect objects, people, etc against which action can then be taken. In some cases, quick action is required particularly in public protection and disaster relief (PPDR) situations. In such cases, the network must provide a) high bandwidth to allow multiple, high-definition video streams to transmit high volumes of data to the recognition system, and b) low latency to ensure a rapid response.

In this experiment, we have investigated not only the use of the 5G-VINNI network to support eMBB requirements, but also the use of edge computing to demonstrate the impact of camera data processed locally, this reducing the network transit time. Wearable cameras provide 5G-connected video content which is streamed through the network to the recognition compute resource. Given there is a trade-off between edge and cloud deployments, the experiment allows us to investigate the optimum compute footprint, where that should be deployed in the 5G-VINNI network, and the dimensions of the network itself to simultaneously support both high bandwidth and low latency.

The experimentation outlined above covers use cases in a variety of verticals and calls upon different 5G requirements. These are summarised in TABLE I.

TABLE I. 5G VINNI TRIALS AT ADASTRAL PARK

| Vertical | Experimentation | 5G capability | User requirement |
|---|---|---|---|
| Industry | Remote robotic control & VR-based immersive app | eMBB/ URLLC | Indoor coverage, Reliability, Security |
| Gaming | Cloud-based gaming | eMBB | Indoor coverage, Latency, QoE |
| Media | QoS-assured connectivity for media broadcast | eMBB/ customised | Indoor coverage, Reliability |
| Health | Connected care for assisted living | eMBB/ URLLC | Indoor & outdoor coverage, Privacy, Reliability, QoE |
| PPDR | Situation recognition for public safety | eMBB/ URLLC | Indoor & outdoor coverage, Reliability, Latency |

## V. LESSONS LEARNED

The 5G-VINNI project has provided, and will continue to provide, valuable learning in the design, installation, operation and experimentation of 5G networks.

### A. Design and installation

In designing the 5G-VINNI UK network, Samsung has employed commercial network equipment. Radio hardware and all software is the equivalent implementation to that used in the commercial network implementation employed in one of the three Korean operator implementations for which Samsung is an equipment supplier. This meant that some key learnings that may have been achieved in the design phase were not possible. Scalability, integration, interface optionality and network topology were all removed as possible topics since the design had to follow that prescribed by the product. Equally however, this offered benefits in the implementation phase since large amounts of integration and testing could be skipped – the implementation of the network, in large part, worked immediately.

In planning coverage for mmWave technology, beam forming, reflected paths and obstacles needed to be taken in to account. This meant that a highly accurate and specialised radio planning tool was needed, and far more detailed information regarding the environment where installation was to take place was required. Not only do buildings, trees and flat surfaces have an impact, but specific details such as window locations on buildings also have an impact. This places a strong focus on accurate survey and map details which was performed using Samsung's in house radio planning tool.

The variance in size of antenna units had two impacts. Structural planning for 3.6 GHz units was based on more traditional antenna installation expectations for weight and surface area (and resulting wind shear), and while the installed units are similar in size to 4G antenna units, when higher order MIMO, and/or higher numbers of antenna elements for higher Tx/Rx, wind shear may become a significant issue in locating antenna units safely.

Converse to this, mmWave antenna units are light and versatile. Weighing under 10kg and being smaller than 10 litres in volume, the planning for suitable locations allowed a range of options including lamppost mounting or simple attachment directly on to buildings, to be considered.

Since all Network Functions for Core for the DU and CU elements of the RAN were softwarised, the functional network design was mostly about software architecture. The planning for physical installation sites and servers were independent of VNF's instantiation. We experienced this during the design phase when a decision was taken to swap server vendor – this had no impact on the functional design of the network, demonstrating that the VNF's being used were entirely versatile and implementable on any suitable server.

As a consequence of subsequent operation and the undertaking of test and experimentation activities, a key lesson for the design phase is to ensure that availability of compute resource (cores and memory) for experimenters, or in a commercial network, third party partner applications and VNFs, is planned for. The extent to which additional resource is needed would depend upon each Network Operator's intended third-party hosting strategy.

Installation of the network was faster than expected – the project plan allowed six weeks for VNF instantiation and testing, but this phase was completed in three weeks. This was in part as a result of the VNF software loads being taken from a commercial Samsung 5G network deployment, and hence requiring little integration. In addition, implementation of VNFs via NFV Orchestrator and Virtualised Infrastructure Manager (VIM) meant network topology discovery happened automatically during provisioning.

### B. Operation

The 5G-VINNI infrastructure at Adastral Park is the first of its kind on this site and therefore most of its day-to-day operation experiences provides useful learning in particular:

- Coverage: Despite undertaking some initial analysis as part of the design, we have seen some unusual coverage patterns in certain parts of the site. Given the site is a typical multi-building business park we were expecting to see a fair number of low coverage areas due to building shadowing. However, our experience confirms that scattering and reflections improve coverage in many of these 'low coverage' areas. Work is now underway to investigate the use of reflective surfaces to boost the signal in these low coverage areas.

- mmWave system: For access to the mmWave radio system we used a set of indoor and outdoor CPEs. The outdoor CPE was designed to be fixed to a building structure. However, the indoor CPE was a portable device about the size of a small WiFi router and designed to be placed in a window facing the mmWave antenna. This allowed us to quickly test indoor coverage levels in different locations in the many different buildings on the site. Furthermore, with the use of a portable power supply, we ran drive tests around the Adastral Park roadways. This has provided indications to help with any connected vehicle applications we may wish to test in future applications.

- 5G UE availability: There has been a high demand for access to the radio systems, particularly the 3.6 GHz NR. Our experience with previous experimental radio testbeds is that, at least in their early incarnations, access would often be via a custom-built UE, often provided in limited numbers. We have been fortunate that the UE used for access to the NR in 5G-VINNI is the Samsung Galaxy S10, available as an off-the-shelf commercial product. This has allowed us to keep pace with the demand for access to the testbed.

- Administration: We have a process in place to provide first and second-line support for the testbed, by BT and Samsung, respectively. The initial intention was for the process to support early back-off to second line support for technical matters. However, particularly in the early period of testbed operation, we experienced this to be sub-optimal in terms of response to faults, re-configurations, etc. Therefore, we have revised our support processes for BT staff have full access to the administrative interfaces of the radio systems to fix faults and make system changes.

*C. Experimentation*

As outlined in previous sections, the purpose of the testbed is to provide an open platform for experimentation. Making the testbed available to BT and Samsung's partners has had a number of consequences:

- *Interacting with experimentations:* Many of our partners have developed applications that they have deployed in a 4G environment and they have been keen to see what 5G can offer. We have seen very high demand for testbed usage and as a result developed a process which we have used with experimenters to assess their interest and use of the testbed. This has also enabled us to spread the load on both the use of the infrastructure (particularly UEs and CPEs) and the time of our testbed professionals. More recently, this local process has been used as the basis for a more formalised on-boarding process to be used for all verticals across the 5G-VINNI project [6]. Our initial expectation was that organisations would come to the site, integrate with the network and test their application. As the integration and test phase has required many weeks' worth of activity, we have designed a dedicated 5G-VINNI co-development digital laboratory which organisations can access to test their solution (as described in Section IV) on the testbed. This 'co-dev lab' provides a secure and exclusive space for testing and has been designed to make use of digital media to allow partner organisations to advertise their brands, and thus aid with their public relations.

- Response to use case requirements: Our initial thoughts as to use cases that we might demonstrate on the testbed in the initial months of operation were aimed at public protection & disaster relief and vehicle connectivity. While experimentation in these areas is currently being developed, we have seen a widely different set of applications being deployed, as outlined in Section IV. One consequence of this is a request to make adjustments to the orientation of the 3.6 GHz and LTE antenna systems to provide improved coverage in an area of the site suitable for drone flying.

VI. SUMMARY

In this paper, we have presented our experience of building a 5G testbed platform in the UK that enables verticals to test their deployments and share our 5G industry outlook. The context for our testbed, in the light of 5G evolution, has been described, as well as an overview of the architecture and physical design of the UK facility.

We have provided a description of the test platform set-up and have summarised a set of use cases that we have tested across five major industry verticals – industry, gaming, media, health and PPDR. Finally, we have reported our experience of design, installation, operation and experimentation in the form of a set of 'lessons learned'.

The 5G-VINNI testbed continues to evolve with the move to 5G stand-alone core planned for later this year. In addition, we have an increasing pipeline of new 5G innovations to experiment with and demonstrate in our dedicated co-development digital laboratory.

ACKNOWLEDGEMENT

This work is fully funded by 5G-VINNI EU H2020 research and innovation programme under grant agreement No 815279.